\documentclass[preprint, 11pt,english]{aastex}
\usepackage{natbib}
\usepackage{graphicx}
\usepackage{babel}
\usepackage{times}
\begin{document}
\title{Coronal Loop Expansion Properties Explained Using Separators}
\author{Joseph E. Plowman, Charles C. Kankelborg, Dana W. Longcope}
\affil{Department of Physics \\ Montana State University \\ Bozeman, MT 59717}

\begin{abstract}
    One puzzling observed property of coronal loops is that they are of roughly constant thickness along their length. Various studies have found no consistent pattern of width variation along the length of loops observed by TRACE and SOHO. This is at odds with expectations of magnetic flux tube expansion properties, which suggests that loops are widest at their tops, and significantly narrower at their footpoints. Coronal loops correspond to areas of the solar corona which have been preferentially heated by some process, so this observed property might be connected to the mechanisms that heat the corona. One means of energy deposition is magnetic reconnection, which occurs along field lines called {\em separators}. These field lines begin and end on magnetic null points, and loops forming near them can therefore be relatively wide at their bases. Thus, coronal energization by magnetic reconnection may replicate the puzzling expansion properties  observed in coronal loops. We present results of a Monte Carlo survey of separator field line expansion properties, comparing them to the observed properties of coronal loops. 
\end{abstract}

\maketitle

\section{Introduction}

Coronal loops are bright EUV and X-Ray plasma structures constituting the ``one-dimensional atmospheres" of which the solar corona is composed \citep{RTV78}. Each of these thin loops is believed to trace out a bundle of magnetic flux. Their cross-sectional areas, $A$, are therefore inversely proportional to the local field strengths, $B$, so that the resulting flux, $\Phi=BA$, is constant.

In both potential and force free models of the solar corona, field strength decreases with height, leading to significant expansion of the loop with respect to the footpoints. Though we lack direct measurements of loop cross-section, it is reasonable, on average, to expect $A$ to scale with the square of the sky plane projected loop width, $d$. To the extent that coronal loop length and height are correlated, a correlation is also expected between loop length and the degree of expansion from footpoint to loop top.

These simple predictions are not borne out by observation. \cite{klimchuk2000} and \cite{watkoklimchuk2000} measured the ratios of footpoint to midpoint width (called {\em expansion factors}) for coronal loops observed in soft X-Rays and EUV, and found them to be near unity. Moreover, Watko and Klimchuk found no statistically significant correlation between loop length and expansion factor. These results were reinforced by \cite{LFKD06}, who compared observed loops with potential and linear force free extrapolations from the corresponding photospheric fields.

Many explanations have been proposed for the lack of observed expansion in coronal loops. \cite{LFKD06} proposed that tangled or braided magnetic fields might constrain the expansion of the field with height. In the nanoflare model of \cite{Parkernflare88}, braiding leads to heating of the corona by reconnection, which could explain why the loop structures are brighter than the background. \cite{KlimchukDevore07} examined the role of twist in a nonlinear force free model, but were unable to reproduce the low measured expansion factors of coronal loops. \cite{DeForest07} proposed that the loops may be narrower than they appear, and that a combination of telescope PSF, solar background and noise obscure the expansion of the loops. This proposal has initiated an extended debate at conferences and in published literature. \cite{LFDK08} have provided a careful error analysis, concluding that the loop width measurements are robust.

This paper takes an alternative approach to the constant width loop problem. Coronal loops are brighter than the rest of the corona, so there must be some mechanism energizing them. Such a mechanism is likely to prefer some field lines over others. If the corona is heated by reconnection, then the heating, enhanced emission and coronal loops should be associated with magnetic separators (\cite{Longcope96}, for instance). Since separators have magnetic nulls at both ends, we hypothesize that the field on or near a separator would not decrease much with height. Separator loops would therefore have more uniform cross-section than general field lines. 

We test this hypothesis with a Monte Carlo survey of simulated coronal field configurations. Rather than constructing detailed simulations of active regions, we employ an ensemble of simpler randomly generated potential field models, comparing the expansion properties of generic separators to those of representative randomly selected field lines. Section \ref{simulation} describes the model that generated both the separator loops and a control population of random potential loops. The distribution of properties of the resulting loops, including length, height and expansion factor, are analyzed in Section \ref{results}. In Section \ref{conclusion} we conclude that separator field lines correspond closely to the observed properties of coronal loops. 

\section{Numerical Simulation}\label{simulation}
\subsection{Charge Distribution Generation:}\label{sub:CDGen}
In this work, simulated coronal fields are generated by thin flux tubes emerging from the photosphere, while the corona itself is current-free. Such a configuration of flux tubes is well described by a planar arrangement of point sources, so the resulting coronal field is given by Coulomb's law:

\begin{equation}\label{coulomb}
{\bf B}({\bf r})=\sum _{i=1}^{N}q_{i}\frac{{\bf r}-{\bf r}_{i}}{\left|{\bf r}-{\bf r}_{i}\right|^{3}}
\end{equation}

In order to assess the properties of coronal fields under this approximation, we generated many such charge distributions with randomized charge parameters:
\begin{itemize}
\item The total number of charges in a distribution, $N$, was varied uniformly between 10 and 20. 
\item The charge positions, ${\bf r}_i$, were uniformly distributed over a square plane.
\item The charges $q_i$ on each monopole were drawn from an exponential distribution, with randomly chosen sign. The total charge was then set to zero by rescaling all charges of the polarity with less total flux.
\end{itemize}

\begin{figure}
\begin{center}\includegraphics[width=1.0\textwidth]{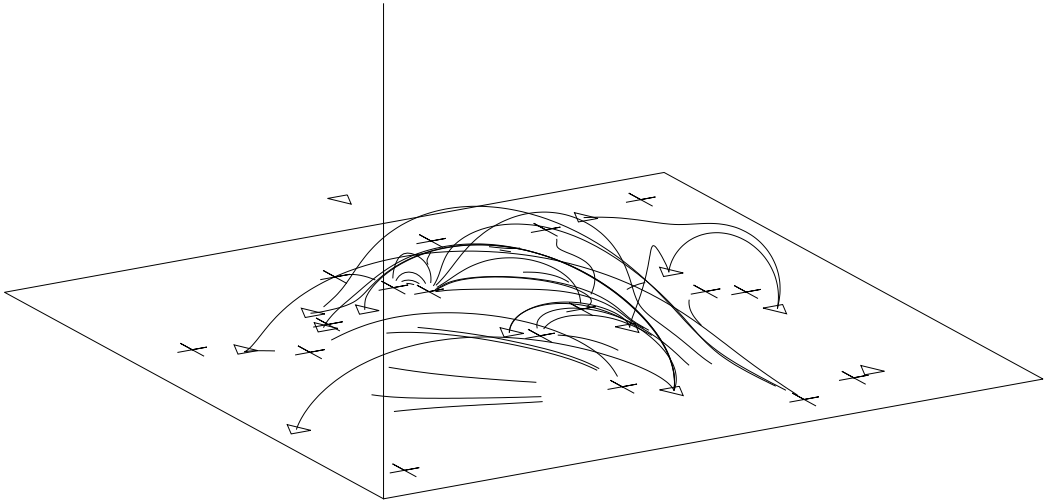}\end{center}
\caption{An example charge configuration. Point charges are represented by
plus signs and separators begin and end on null points (represented
by triangles). All other field lines were randomly chosen.\label{fig1}}
\end{figure}

For each of these charge configurations, we found magnetic null points and traced the separator field lines connecting them, using algorithms described in \cite{Longcope96}. Magnetic field strength and arc length as functions of position were recorded for each separator field line. On average, we found about 3 separators per charge configuration and have data for 1250 separators in all.

Is $\sim 3$ separators per charge configuration reasonable, given our simulation? We tested this result using the relation \citep{CBDL05} $N_S \approx N_D - N + 1$, between the numbers of separators ($N_S$),  flux domains ($N_D$), and flux sources ($N$) in a configuration. The average number of connections between flux sources in a charge configuration (typically equal to the average number of flux domains) was determined by tracing a number of random field lines for each source and tracking which other sources they connected it to. The average number of connections per configuration was found to be $N_D \sim 20$, with an average number of sources per configuration of $N \sim 15$. The average number of separators per configuration ought then to be $N_S \sim 6$, about double the number found. This result is reasonable given known inefficiencies in the algorithms used to find the separators.

This raw data set was further processed to make it more comparable to coronal loop observations. The loop footpoints were defined to be at a height $z_{ct}$, which was at $5 \%$ of the total dimensions of the distribution and $\sim 30 \%$ of the height of a typical separator. To facilitate comparisons between loops, linear interpolation was used to resample the magnetic field strength and height for 500 evenly spaced points along each field line. Also recorded was the total arclength, $L$, of the loop segment lying above $z_{ct}$. At each point on the loop, the approximation $d \propto 1/\sqrt{B}$ was used to relate the magnitude of the magnetic field, $B$, to the loop diameter $d$. Left and right expansion ratios, $R_l$ and $R_r$, were computed for each field line. These were defined as the ratios of the mean loop diameter $d$ on the central 15\% of the field  line to the mean width of the 15\% nearest its left and right footpoints, respectively (Compare \cite{watkoklimchuk2000}).

We found that many of the separators observed had significantly asymmetric field strengths. In order to better understand the effects of this, we arranged all of the field lines so that the side with higher field (on average) was always on the right. Thus, our field lines are systematically wider on their left sides than on their right, and the resulting left side expansion ratios are smaller than the right side expansion ratios.

\subsection{Random field line generation}
We also generated a representative `control' sample from the overall population of field lines. The most obvious choice for a representative population of field lines would give equal weight to equal flux. Such a `flux-weighted' population of field lines may be generated by picking a source, weighted by the magnitude of its charge, stepping a short distance away from the charge in a random direction, and then tracing the field from that point. 

The flux-weighted approach was rejected, however, because such field lines tended not to extend significantly above the charge plane. This is unsurprising, since the fields will be strongest close to the plane. We therefore note that most field lines (weighted by flux) do not resemble coronal loops.

In light of this, we generated a new population
of random field lines whose heights were picked based on a histogram
of the separator heights. This was accomplished by choosing random initial points ${\bf r}_{0}$ from a distribution uniform over the field region, tracing the field lines in
both directions and then discarding members of the resulting population until the final set had roughly the same height distribution as the previously generated population of separators. 

For each of these random field lines, we recorded the same field line properties as for the separators. As with the separator field lines, the random field lines were arranged so that the side with higher field was on the right.

\section{Simulation Results}\label{results}
\begin{figure}
\begin{center}\includegraphics[width=1.0\textwidth]{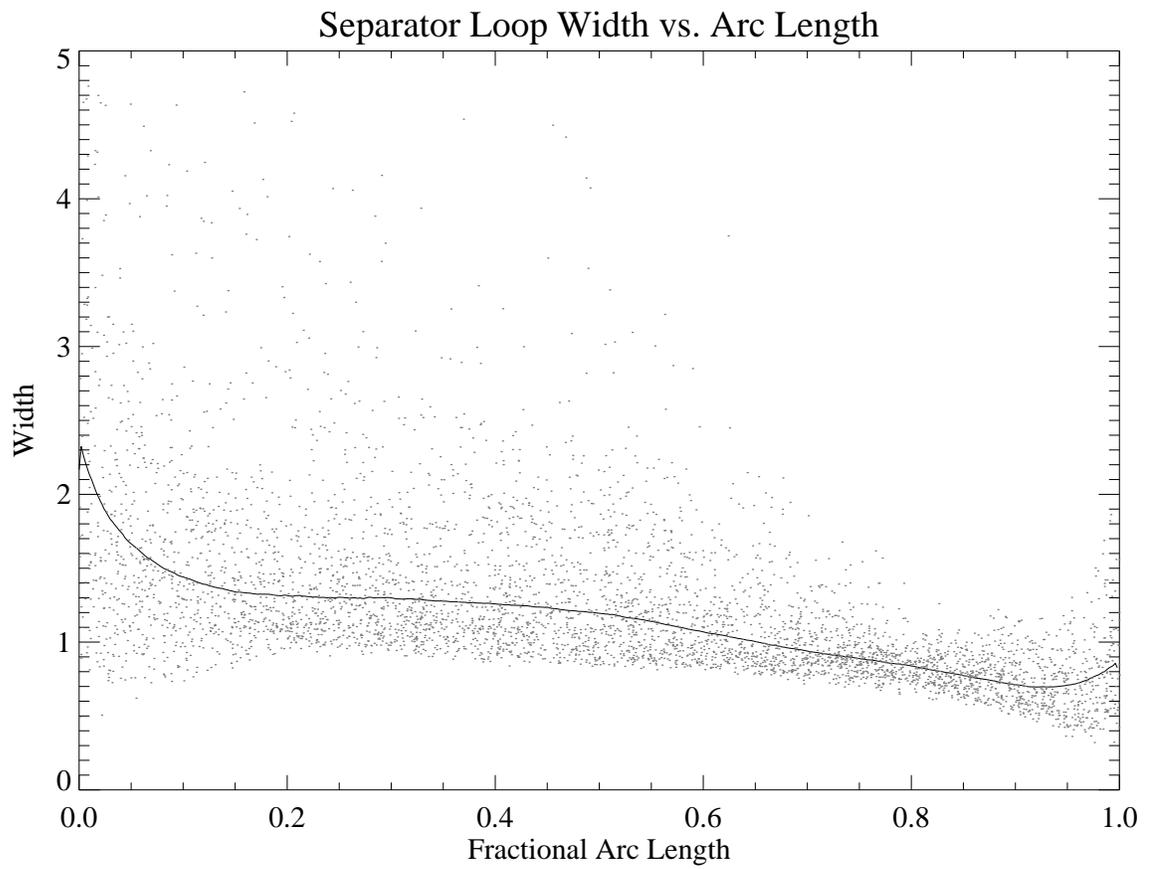}\end{center}
\caption{Separator loop width as function of fractional arc length. Each loop scaled so that the mean width is unity. Solid line is the median width at that length coordinate.\label{mediansepwidths}}
\end{figure}

\begin{figure}
\begin{center}\includegraphics[width=1.0\textwidth]{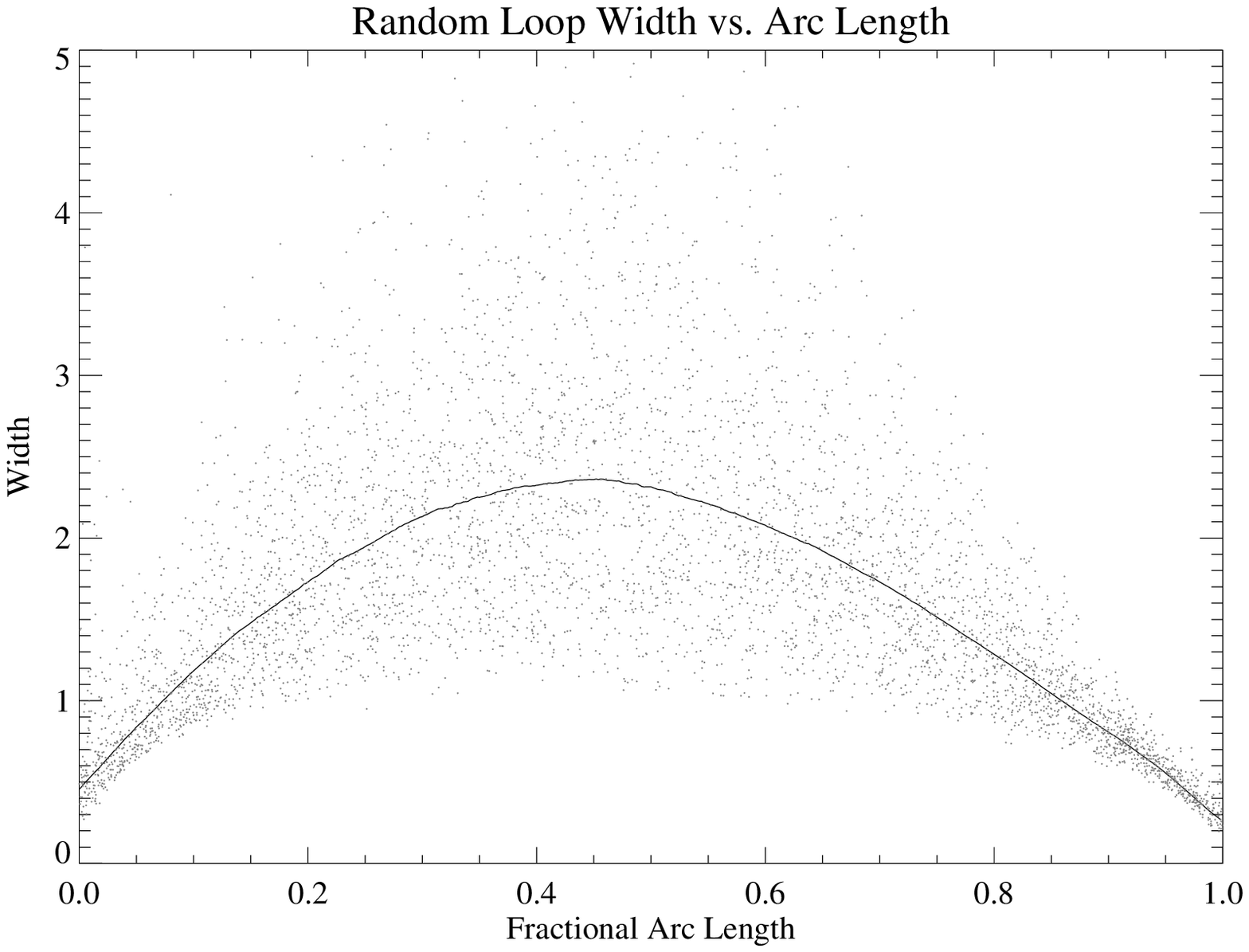}\end{center}
\caption{Random loop width as function of fractional arc length. Each loop is scaled so that the mean width is unity. Solid line is the median width at that length coordinate.\label{medianranwidths}}
\end{figure}

To give an idea of the typical width profile of these field lines, figures \ref{mediansepwidths} and \ref{medianranwidths} show loop widths as a function of fractional arc length for separator and random field lines, respectively. The median curves in these plots (i.e., the solid lines) show that the separator field lines tend to have a much smaller range of widths than the random field lines, and that the separators have some significant tendency towards skewness (one side of the field line is significantly wider than the other, as discussed in Section \ref{sub:CDGen}), while the random field lines do not. The median separator width plot has an intriguing resemblance to Fig. 9 of \cite{watkoklimchuk2000}.

\subsection{Expansion Ratios}
\begin{figure}
\begin{center}\includegraphics[width=1.0\textwidth]{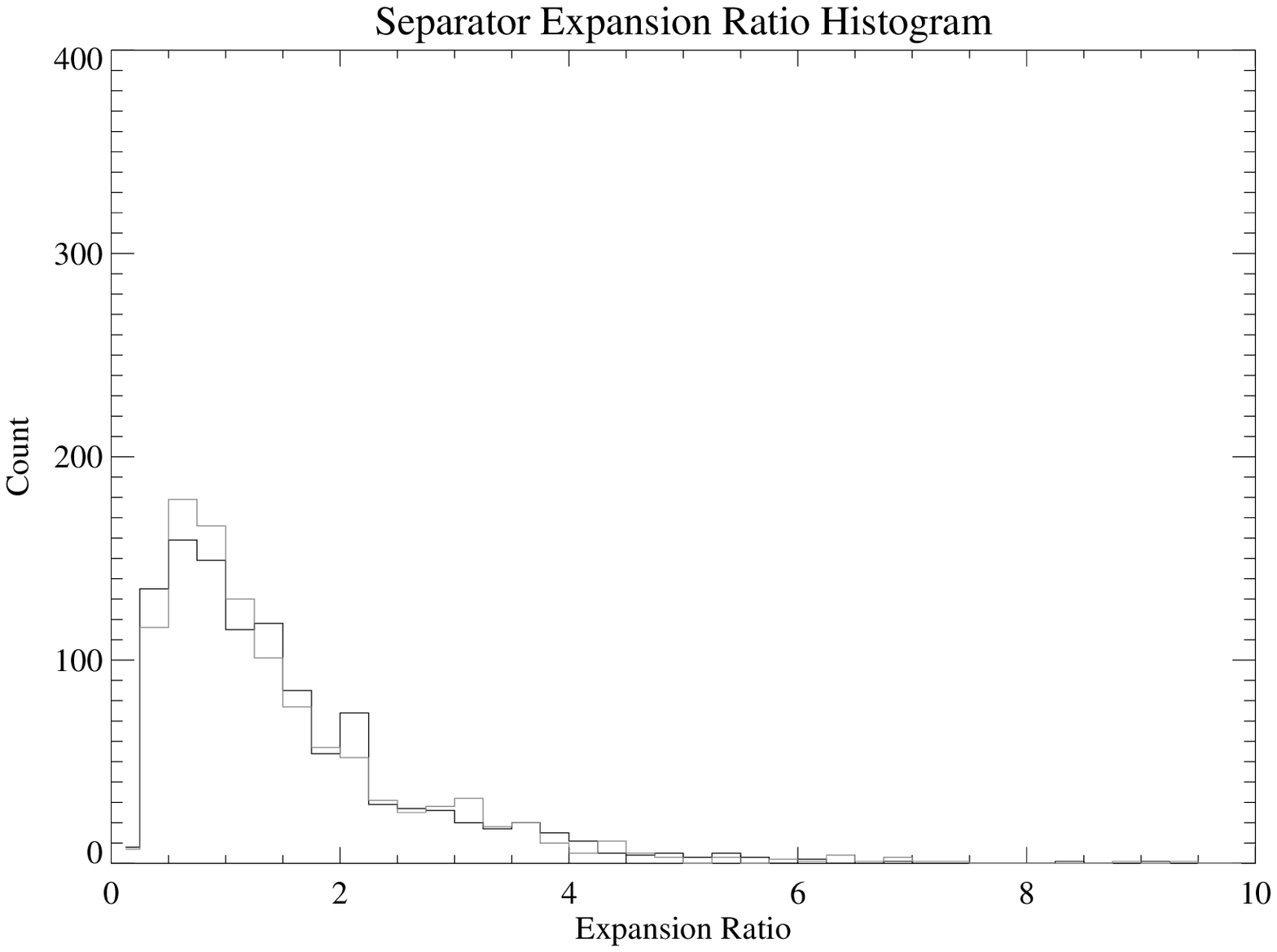}\end{center}
\caption{Expansion ratio histograms for separator loops. Left expansion ratios are in black, right expansion ratios in gray.\label{sepexpratios}}
\end{figure}

\begin{figure}
\begin{center}\includegraphics[width=1.0\textwidth]{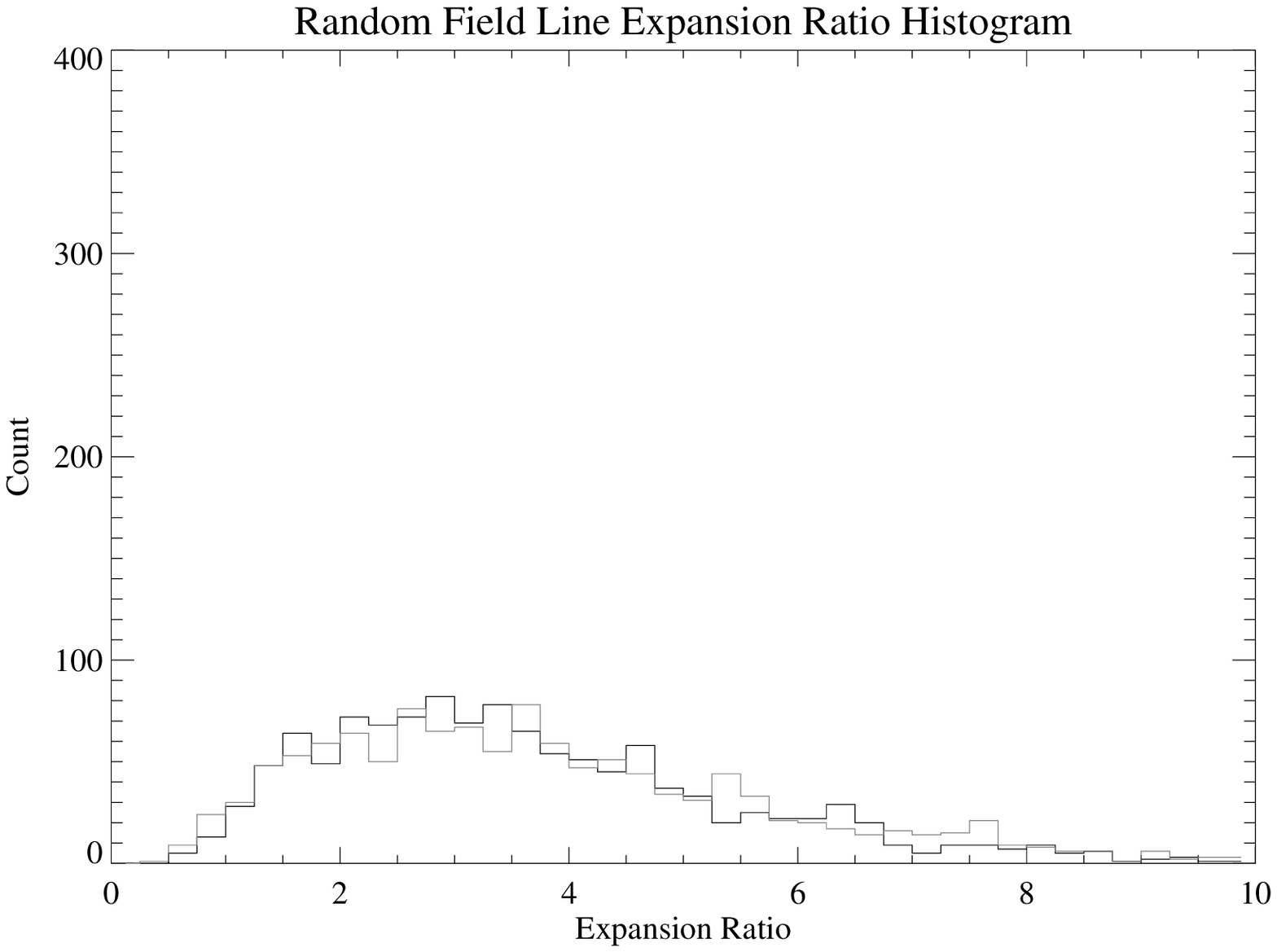}\end{center}
\caption{Expansion ratio histograms for random loops. Left expansion ratios are in black, right expansion ratios in gray.\label{ranexpratios}}
\end{figure}

The separators generally have much lower expansion ratios than the random field lines. Median expansion ratio for the left and right separators were $0.69$ and $1.65$, respectively, while those for the random field lines were $2.74$ and $4.35$. Histograms of these expansion ratios are shown in figures \ref{sepexpratios} and \ref{ranexpratios}. The expansion factors divide the separators and the random field lines into distinct populations, with random field lines having expansion ratios appreciably greater than one, while the separator expansion ratios are clustered around unity. By this measure, the separator field lines resemble the coronal loops observed by \cite{watkoklimchuk2000}, while the random field lines are distinctly different.

\cite{watkoklimchuk2000} compare expansion ratio to length for 15 non-flare
coronal loops, and find no statistically significant correlation. We compared our simulated data sets with these findings by checking for correlation of expansion ratio with loop length, for both separator and random field lines. 

\begin{figure}
\begin{center}\includegraphics[width=1.0\textwidth]{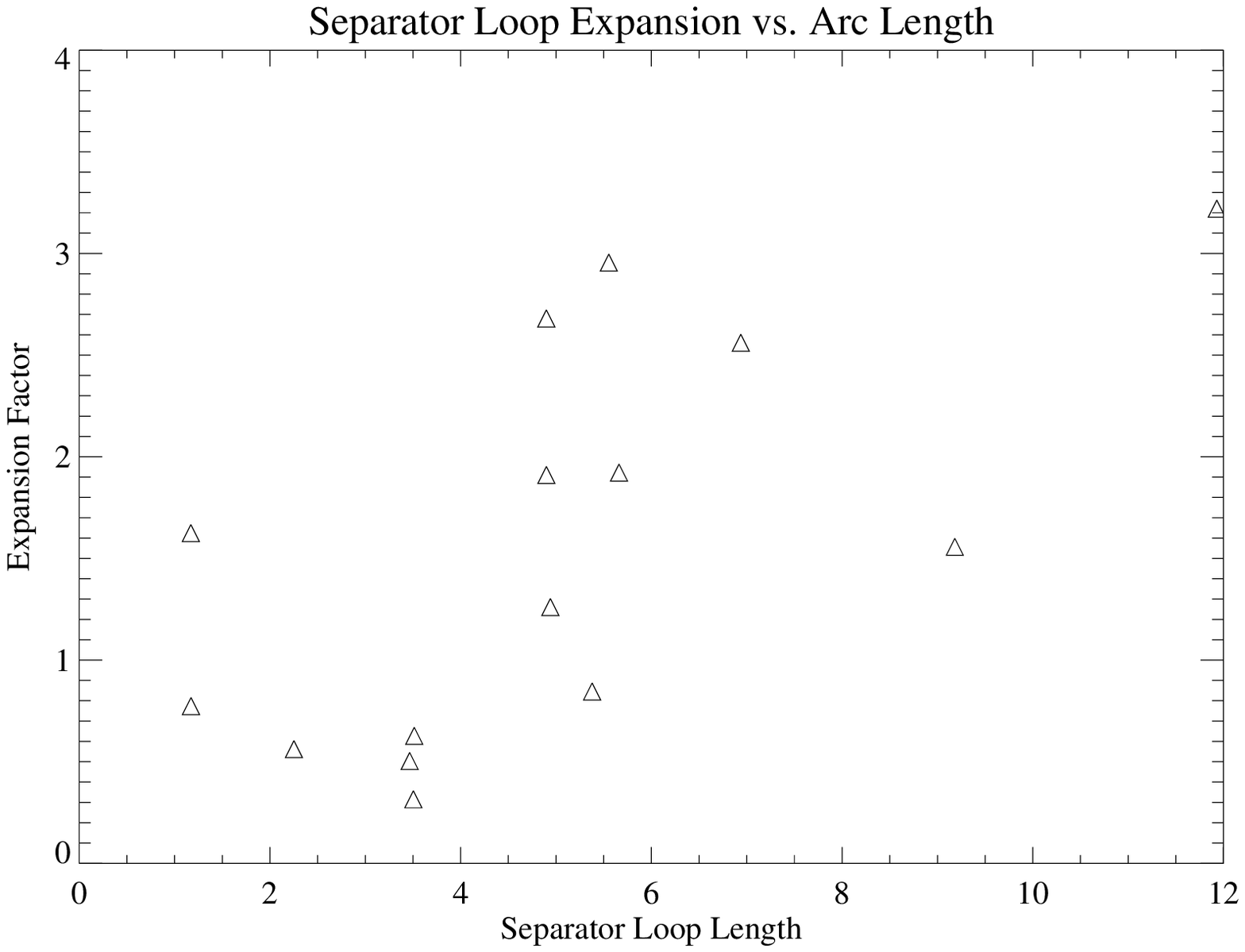}\end{center}
\caption{Plot of expansion factor versus loop length for 15 separator loops. \label{sep15expratios}}
\end{figure}

\begin{figure}
\begin{center}\includegraphics[width=1.0\textwidth]{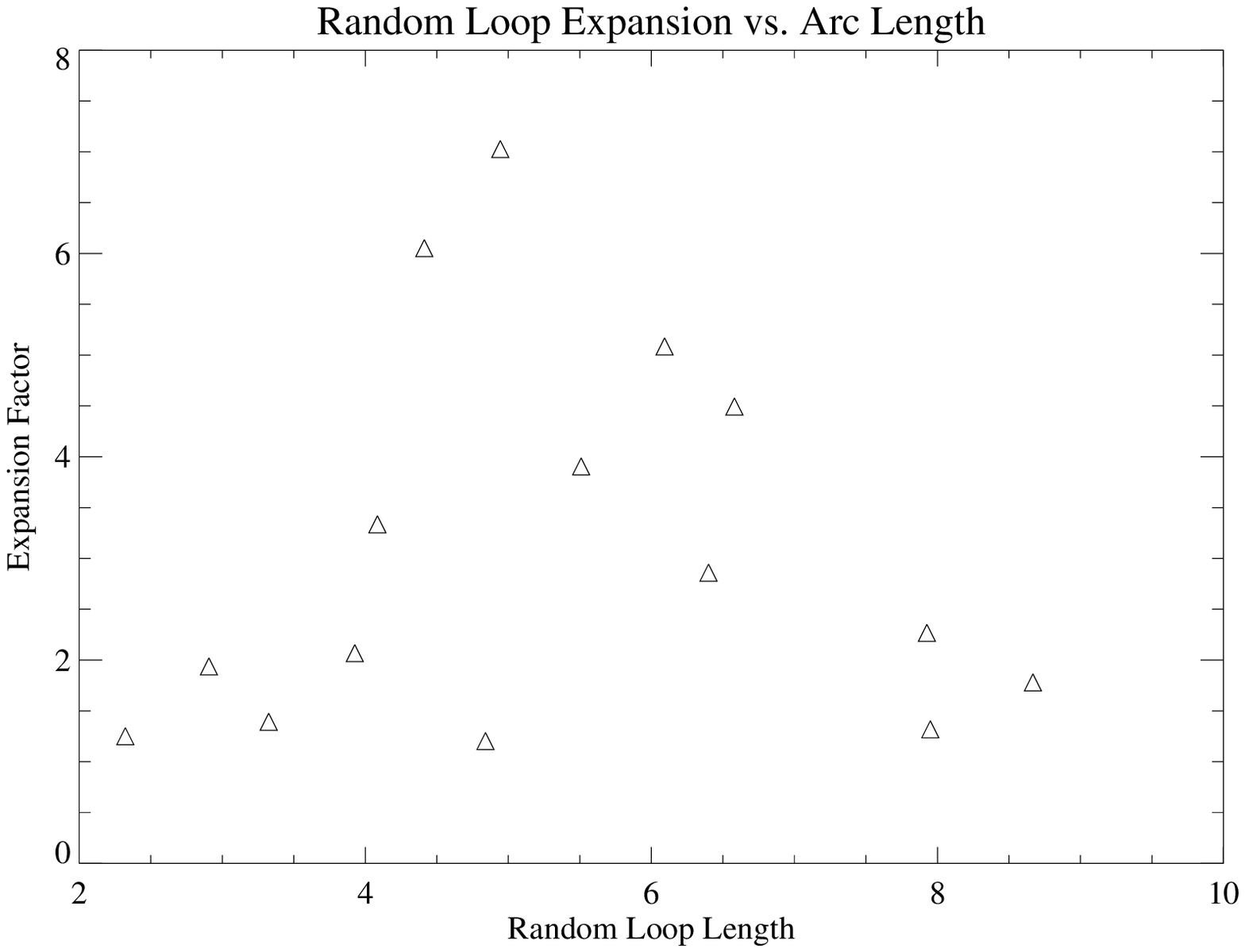}\end{center}
\caption{Plot of expansion factor versus loop length for 15 random loops. \label{ran15expratios}}
\end{figure}

In order to make our analysis correspond more closely with small data set analyses such as \cite{watkoklimchuk2000}, we operated on sets of 15 field lines drawn at random from the overall population (figures \ref{sep15expratios} and \ref{ran15expratios} show example draws for separator and random field lines, respectively). With each set of field lines, we used
a Kendall's tau statistic to check for correlation of expansion ratio (Either left or right, chosen at random for each field line) with loop length. This was repeated 5000 times to make full use of the data set.

In both the random and separator field line cases, the analysis revealed a weak correlation between expansion ratios and loop lengths. The median of the tau statistics for separators was $0.25$, while that for random field lines was $0.33$. These tau values correspond, respectively, to an $82\%$ and $92\%$ confidence level. Lack of correlation like that observed observed by \cite{watkoklimchuk2000} is therefore consistent with the separator loops in our data set, but less consistent with the random loops.

What are we to make of these correlations, or lack thereof? In the case of the separator loops, lack of expansion is expected, as they are wide at both midpoint and footpoint. But why is the correlation also weak for our random loops? Recall that expectations of correlation between expansion and length are based on the following assumptions:

\begin{enumerate}
\item Field strength falls off with height, and has an inverse relationship with the width along the loop. Therefore, taller loops should have more expansion.

\item Inspection of coronal loops suggests that loop length scales with loop height.

\item If expansion is correlated with height and height is correlated with length, then expansion should be correlated with length.

\end{enumerate}

\begin{figure}
\begin{center}\includegraphics[width=1.0\textwidth]{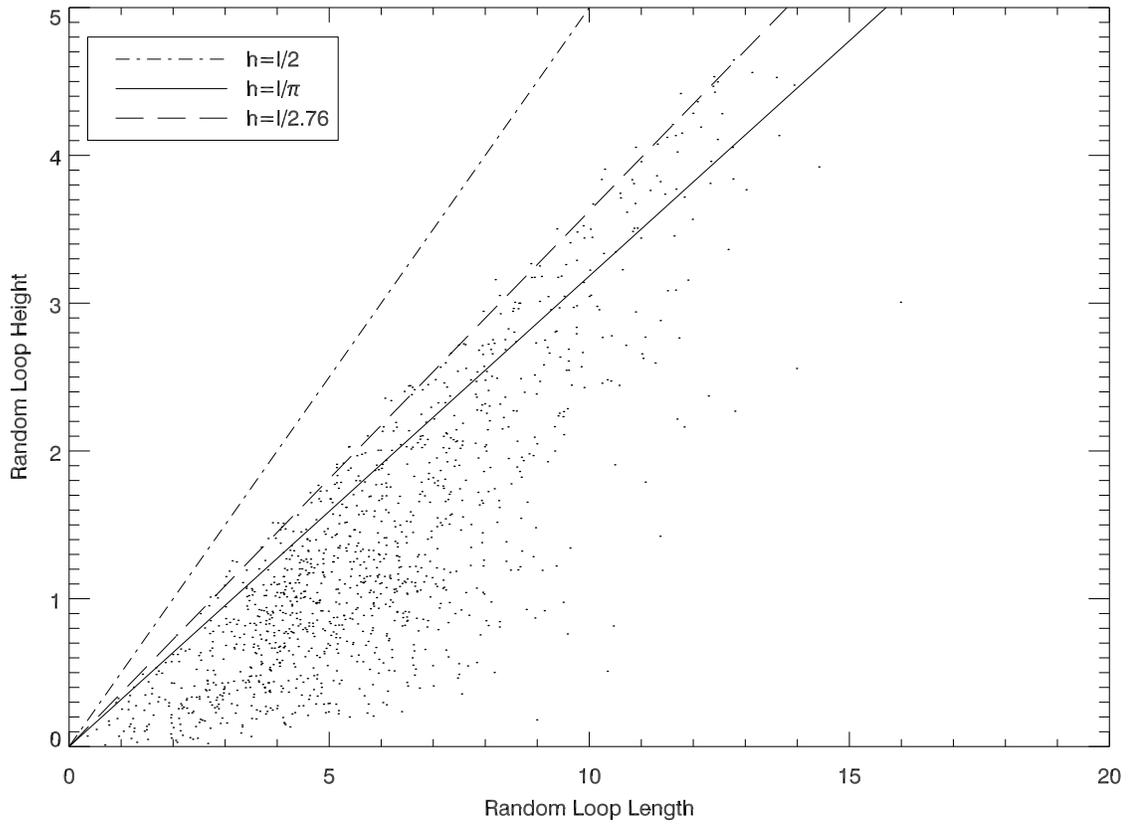}\end{center}
\caption{Scatter plot of loop length versus loop height for all random field loops. Included for reference are lines corresponding to the tallest possible loop of length $l$, $h=l/2$, to a semicircular arc, $h=l/{\pi}$, and to the tallest circular arc of length $l$, $h\approx l/2.76$.\label{ranhvsl}}
\end{figure}

Inspection of our simulated data set suggests that the second of these assumptions is only weakly satisfied by random loops. While there is a correlation between loop length and height, and the median Kendall's tau for 15-element data sets is $0.62$ ($\sim 99.9\%$ significance), visual inspection of the data set suggests (see fig. \ref{ranhvsl}) that it is due largely to the obvious geometrical constraint that the loop length must be at least twice the loop height. These random field line data fill the parameter space almost uniformly below $h = l/2.76$, which corresponds to the tallest possible circular arc. The first of the above assumptions was also found to be weakly satisfied by the random loops, with the 15-sample Kendall's taus for correlation between expansion factor and height having a median value of $0.50$ (also $\sim 99\%$ significance).

\begin{figure}
\begin{center}\includegraphics[width=1.0\textwidth]{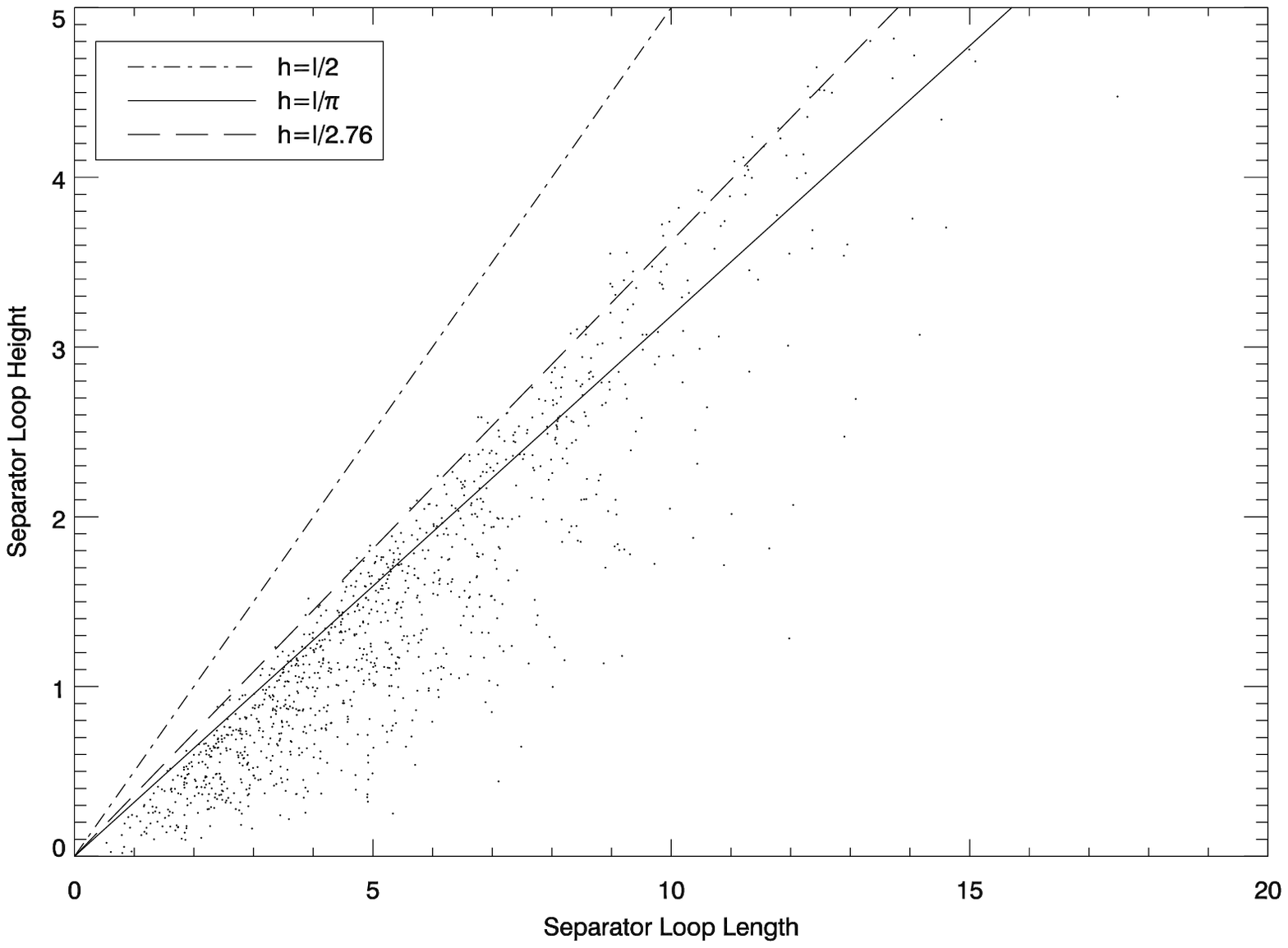}\end{center}
\caption{Scatter plot of loop length versus loop height for all separator field lines. Included for reference are lines corresponding to the tallest possible loop of length $l$, $h=l/2$, to a semicircular arc, $h=l/{\pi}$, and to the tallest circular arc of length $l$, $h\approx l/2.76$.\label{sephvsl}}
\end{figure}

When the same properties were checked for separator loops, height was not found to be significantly correlated with expansion factor: The median 15-sample Kendall's tau was 0.20, corresponding to a $74\%$ significance. This is as expected, given the separator loop's relatively small footpoint diameter. Loop length scales very nicely with height, however, with a linear trend that is readily apparent in a scatter plot of the whole data set (fig. \ref{sephvsl}), and 15-sample median Kendall's taus of $0.77$ ($99.994\%$ probability of correlation). Thus, the separators are consistent with the observationally motivated intuition relating loop length and height, in addition to being consistent with the observed low expansion factors and lack of correlation between expansion and loop length.

Mathematically, the third of the above assumptions is only true in the presence of `strong' (Kendall's tau significantly greater than $\sim 0.5$) correlations. Mere statistical significance, in the sense of falsifying the null hypothesis (that expansion is independent of height, and that height is independent of length), is not sufficient to ensure a correlation between between expansion and length. In the case of the random loops, the link connecting expansion with height has somewhat weak correlation, so the resulting weak correlation between expansion and length is unsurprising. In the case of the separators, even though the connection between length and height is quite strong, the connection between height and expansion is very weak, and therefore there is no significant connection between length and expansion.

\section{Conclusion}\label{conclusion}
This potential field study demonstrates that separator loops possess expansion factors near unity (no significant expansion), consistent with observations of the solar corona. Like coronal loops, the separators stand tall, with a strong correlation between arc length and height. Random loops chosen purely on the basis of flux weighting usually do not extend to any significant height above their footpoints. This indicates that coronal loops are not representative of the overall population of coronal field lines. A second random control population, with heights constrained to more closely resemble coronal loops, had typical expansion factors between 3 and 4, also inconsistent with observed coronal loop expansion properties.

We found that both loop populations were consistent with the lack of correlation between expansion factor and loop width observed by \cite{watkoklimchuk2000}. While there was no statistically significant correlation between expansion ratio and loop length for small 15 element samples such as those observed by \cite{watkoklimchuk2000}, a weak correlation was present for larger sample sizes. This suggests that the lack of correlation observed by \cite{watkoklimchuk2000} and other similar studies may be attributable to the small size of the observational data sets, rather than a lack of any significant correlation in coronal loop populations. Among the randomly chosen loops, the low degree of correlation between length and expansion corresponds a lack of physically significant correlation between length and height. By contrast, the low correlation between length and expansion among separator loops corresponds to the separator loop expansions being narrowly distributed around unity.

\noindent This work was supported in part by NASA grant \# NNX07AI01G.

\bibliographystyle{apj}
\bibliography{apj-jour,paper6}

\begin{thebibliography}{10}
\expandafter\ifx\csname natexlab\endcsname\relax\def\natexlab#1{#1}\fi

\bibitem[{{Beveridge} \& {Longcope}(2005)}]{CBDL05}
{Beveridge}, C. \& {Longcope}, D.~W. 2005, \solphys, 227, 193

\bibitem[{{DeForest}(2007)}]{DeForest07}
{DeForest}, C.~E. 2007, \apj, 661, 532

\bibitem[{{Klimchuk}(2000)}]{klimchuk2000}
{Klimchuk}, J.~A. 2000, \solphys, 193, 53

\bibitem[{{Klimchuk} \& {DeVore}(2007)}]{KlimchukDevore07}
{Klimchuk}, J.~A. \& {DeVore}, C.~R. 2007, in Bulletin of the American
  Astronomical Society, Vol.~38, Bulletin of the American Astronomical Society,
  164--+

\bibitem[{{Longcope}(1996)}]{Longcope96}
{Longcope}, D.~W. 1996, \solphys, 169, 91

\bibitem[{{L{\'o}pez Fuentes} {et~al.}(2008){L{\'o}pez Fuentes},
  {D{\'e}moulin}, \& {Klimchuk}}]{LFDK08}
{L{\'o}pez Fuentes}, M.~C., {D{\'e}moulin}, P., \& {Klimchuk}, J.~A. 2008,
  \apj, 673, 586

\bibitem[{{L{\'o}pez Fuentes} {et~al.}(2006){L{\'o}pez Fuentes}, {Klimchuk}, \&
  {D{\'e}moulin}}]{LFKD06}
{L{\'o}pez Fuentes}, M.~C., {Klimchuk}, J.~A., \& {D{\'e}moulin}, P. 2006,
  \apj, 639, 459

\bibitem[{{Parker}(1988)}]{Parkernflare88}
{Parker}, E.~N. 1988, \apj, 330, 474

\bibitem[{{Rosner} {et~al.}(1978){Rosner}, {Tucker}, \& {Vaiana}}]{RTV78}
{Rosner}, R., {Tucker}, W.~H., \& {Vaiana}, G.~S. 1978, \apj, 220, 643

\bibitem[{{Watko} \& {Klimchuk}(2000)}]{watkoklimchuk2000}
{Watko}, J.~A. \& {Klimchuk}, J.~A. 2000, \solphys, 193, 77

\end{thebibliography}

\end{document}